\documentclass[aps,preprintnumbers,twocolumn]{revtex4}

\usepackage{graphicx}
\usepackage{bm}
\def\gappeq{\mathrel{\rlap {\raise.5ex\hbox{$>$}}
{\lower.5ex\hbox{$\sim$}}}}

\def\lappeq{\mathrel{\rlap{\raise.5ex\hbox{$<$}}
{\lower.5ex\hbox{$\sim$}}}}

\def\ga{\mathrel{\raise.3ex\hbox{$>$\kern-.75em\lower1ex\hbox{$\sim$}}}}
\def\la{\mathrel{\raise.3ex\hbox{$<$\kern-.75em\lower1ex\hbox{$\sim$}}}}
\def\gev{{\rm \, Ge\kern-0.125em V}}
\def\tev{{\rm \, Te\kern-0.125em V}}
\def\beq{\begin{equation}}
\def\eeq{\end{equation}}

\def\mchi{m_{\chi}}

\def\m12{m_{1\!/2}}

\newcommand{\eg}{{\em e.g.}}

\newcommand{\km}{{\rm km}}
\newcommand{\cm}{{\rm cm}}
\newcommand{\yr}{{\rm yr}}
\newcommand{\s}{{\rm s}}
\newcommand{\ethr}{E_{\rm th}}
\newcommand{\eopt}{E_{\rm opt}}

\begin{document}

\preprint{CERN--TH/2001-298,
          MIT--CTP--3200, UCI--TR--2001--33,
          UMN--TH--2031/01, TPI--MINN--01/49, 
          Snowmass P3-30}

\title{Supersymmetric Dark Matter Detection at Post-LEP Benchmark Points}



\author{John Ellis}
\email[]{John.Ellis@cern.ch}
\affiliation{TH Division, CERN, CH--1211 Geneva 23, Switzerland}
\author{Jonathan L.~Feng}
\email[]{jlf@mit.edu}
\affiliation{Center for Theoretical Physics,
             Massachusetts Institute of Technology,
             Cambridge, MA 02139, USA}
\affiliation{Department of Physics and Astronomy, 
             University of California, Irvine, CA 92697, USA}
\author{Andrew Ferstl}
\email[]{andrew.ferstl@winona.msus.edu}
\affiliation{Department of Physics,      
             Winona State University, 
             Winona, MN 55987, USA}
\author{Konstantin T.~Matchev}
\email[]{Konstantin.Matchev@cern.ch}
\affiliation{Theory Division, CERN,
             CH--1211, Geneva 23, Switzerland}
\author{Keith A.~Olive}
\email[]{olive@umn.edu}
\affiliation{Theoretical Physics Institute, School of Physics and Astronomy,
             University of Minnesota, Minneapolis, MN 55455, USA}

\date{November 22, 2001}

\begin{abstract}

We review the prospects for discovering supersymmetric dark matter in a
recently proposed set of post-LEP supersymmetric benchmark scenarios.  We
consider direct detection through spin-independent nuclear scattering, as
well as indirect detection through relic annihilations to neutrinos,
photons, and positrons. We find that several of the benchmark scenarios
offer good prospects for direct detection through spin-independent nuclear
scattering, as well as indirect detection through muons produced by
neutrinos from relic annihilations in the Sun, and photons from
annihilations in the galactic center.

\end{abstract}

\maketitle



A set of benchmark supersymmetric model parameter choices was recently
proposed~\cite{Battaglia:2001zp} with the idea of exploring the
possible phenomenological signatures in different classes of
experiments in a systematic way. The proposed 13 benchmark points
(labelled A-M) were chosen by first implementing the constraints on
the parameter space of the minimal supersymmetric standard model with 
universal input soft supersymmetry-breaking parameters that are
imposed~\cite{EFGO} by previous experiments, and by requiring the 
calculated
supersymmetric relic density to fall within the range $0.1 <
\Omega_\chi h^2 < 0.3$ preferred by astrophysics and cosmology. Four
general regions of cosmologically allowed parameter space were
identified: a `bulk' region at relatively low $m_0$ and $m_{1/2}$
(points B, C, G, I, and L), a `focus-point'
region~\cite{Feng:2000mn,Feng:2000gh} at relatively large $m_0$ (E and
F), a coannihilation `tail' extending out to relatively large
$m_{1/2}$~\cite{EFOSi,glp} (A, D, H, and J), and a possible `funnel'
between the focus-point and coannihilation regions due to rapid
annihilation via direct-channel Higgs boson
poles~\cite{EFGOSi,Lahanas:2001yr} (K and M).

Here we ask whether the
supersymmetric dark matter candidate, the lightest neutralino, can be
observed in experiments that are underway or in preparation.  These
include direct searches~\cite{direct} for the elastic scattering of
astrophysical cold dark matter particles on target nuclei, and
indirect searches~\cite{Feng:2001zu} for particles produced by the
annihilations of supersymmetric relic particles inside the Sun or
Earth, in the galactic center, or in the galactic halo.

It was found previously~\cite{Battaglia:2001zp} that, in $g_\mu -
2$-friendly scenarios, supersymmetry was relatively easy to discover
and study at future colliders such as the LHC and a linear collider
with $E_{CM} = 1$~TeV, which would be able to observe rather
complementary subsets of superparticles. However, some of the other
points might escape detection, except via observations of the lightest
neutral Higgs boson. The most difficult points were typically those in
the focus-point region, at the tip of the coannihilation tail, or
along the rapid-annihilation funnels, with points F, H, and M being
particularly elusive.

In this report, we summarize our results~\cite{newpaper} on the
prospects for the direct and indirect detection of astrophysical dark
matter for each of these benchmark points, taking into account the
sensitivities of present and planned detectors.

\begin{figure}[t]
\includegraphics[height=2.3in]{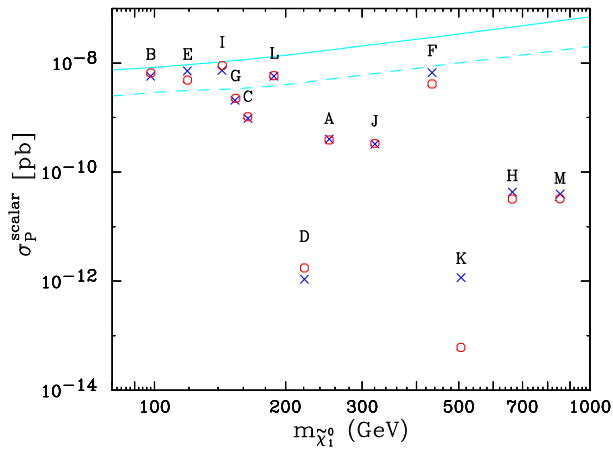}%
\caption{\it Elastic cross sections for spin-independent
neutralino-proton scattering.  The predictions of {\tt SSARD} (blue
crosses) and {\tt Neutdriver} (red circles) are compared.  Projected
sensitivities for CDMS II~\cite{Schnee:1998gf} and
CRESST~\cite{Bravin:1999fc} (solid) and GENIUS~\cite{GENIUS} (dashed)
are also shown.}
\label{fig:direct}
\end{figure}

In Fig.~\ref{fig:direct}, we present the spin-independent
cross-section for neutralino-proton scattering for each benchmark
point using two different codes: {\tt
Neutdriver}~\cite{Jungman:1996df} and {\tt
SSARD}~\cite{ssard}. (Experiments sensitive to spin-dependent
scattering have inferior reach~\cite{newpaper}.)  We find reasonable
agreement, with the largest differences arising for points D and K,
where the cross-section is abnormally small due to
cancellations~\cite{Ellis:2001qm}. For any given $\tan\beta$, the
cancellations occur only for a specific limited range in the
neutralino mass.  Unfortunately, points D and K fall exactly into this
category.

Fig.~\ref{fig:direct} also shows the projected sensitivities for CDMS
II~\cite{Schnee:1998gf}, CRESST~\cite{Bravin:1999fc}, and
GENIUS~\cite{GENIUS}. Comparing the benchmark model predictions with
the projected sensitivities, we see that models I, B, E, L, G, F, and
C offer the best detection prospects. In particular, the first four of
these models would apparently be detectable with the proposed GENIUS
detector.


Indirect dark matter signals arise from enhanced pair annihilation
rates of dark matter particles trapped in the gravitational wells at
the centers of astrophysical bodies. While most annihilation products
are quickly absorbed, neutrinos may propagate for long distances and
be detected near the Earth's surface through their charged-current
conversion to muons.  High-energy muons produced by neutrinos from the
centers of the Sun and Earth are therefore prominent signals for
indirect dark matter detection~\cite{Feng:2001zu,neutrinos}.

\begin{figure}[t]
\includegraphics[height=2.3in]{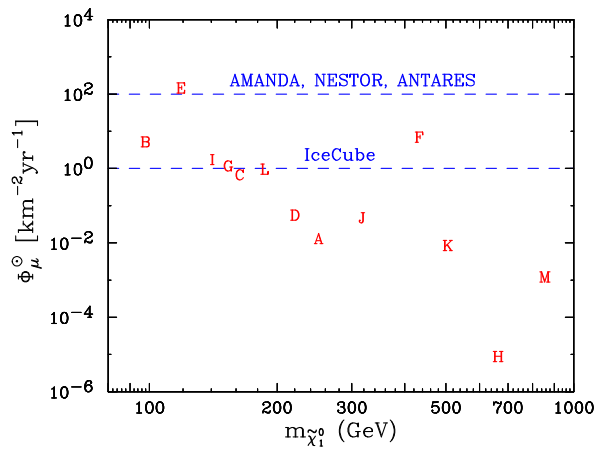}%
\caption{\it Muon fluxes from neutrinos originating from relic
annihilations inside the Sun. Approximate sensitivities of near future
neutrino telescopes ($\Phi_{\mu} = 10^2~\km^{-2}~\yr^{-1}$ for AMANDA
II~\cite{AMANDA}, NESTOR~\cite{NESTOR}, and ANTARES~\cite{ANTARES},
and $\Phi_{\mu} = 1~\km^{-2}~\yr^{-1}$ for IceCube~\cite{IceCube}) are
also indicated.  }
\label{fig:muons}
\end{figure}

Muon fluxes for each of the benchmark points are given in
Fig.~\ref{fig:muons}, using {\tt Neutdriver} with a fixed constant
local density $\rho_0 = 0.3~\gev/\cm^3$ and neutralino velocity
dispersion $\bar{v} = 270~\km/\s$.  For the points considered, rates
from the Sun are far more promising than rates from the Earth.  For
the Sun, muon fluxes are for the most part anti-correlated with
neutralino mass~\cite{newpaper}, with two strong exceptions: the focus
point models E and F have anomalously large fluxes.  In these cases,
the dark matter's Higgsino content, though still small, is
significant, leading to annihilations to gauge boson pairs, hard
neutrinos, and enhanced detection rates.

The potential of current and planned neutrino telescopes has been
reviewed in~\cite{Feng:2001zu}. The exact reach depends on the
salient features of a particular detector, \eg, physical dimensions,
muon energy threshold, etc., and the expected characteristics of the
signal, \eg, angular dispersion, energy spectrum and source (Sun or
Earth).  Two sensitivities, which are roughly indicative of the
potential of upcoming neutrino telescope experiments, are given in
Fig.~\ref{fig:muons}. For focus point model E, where the neutralino is
both light and significantly different from pure Bino-like, detection
in the near future at AMANDA II~\cite{AMANDA}, NESTOR~\cite{NESTOR},
and ANTARES~\cite{ANTARES} is possible.  Point F may be within reach
of IceCube~\cite{IceCube}, as the neutralino's significant Higgsino
component compensates for its large mass.  For point B, and possibly
also points I, G, C, and L, the neutralino is nearly pure Bino, but is
sufficiently light that detection at IceCube may also be possible.

Muon energy thresholds specific to individual detectors have not been
included.  For AMANDA II and, especially, IceCube, these thresholds
may be large, significantly suppressing the muon signal in models with
$\mchi$ less than about 4 to 6 $E_{\mu}^{\rm
th}$~\cite{Bergstrom:1997tp,Barger:2001ur}.  Note also that, for
certain neutralino masses and properties, a population of dark matter
particles in solar system orbits may boost the rates presented here by
up to two orders of magnitude~\cite{Damour:1998rh}.  We have
conservatively neglected this possible enhancement.


As with the centers of the Sun and Earth, the center of the galaxy may
attract a significant overabundance of relic dark matter
particles~\cite{Urban:1992ej}.  Relic pair annihilation at the
galactic center will then produce an excess of photons, which may be
observed in gamma ray detectors.  While monoenergetic signals from
$\chi \chi \rightarrow \gamma \gamma$ and $\chi \chi \rightarrow
\gamma Z$ would be spectacular~\cite{Bergstrom:1998fj}, they are
loop-suppressed and unobservable for these benchmark points. We
therefore consider continuum photon signals here.

\begin{figure}[t]
\includegraphics[height=2.3in]{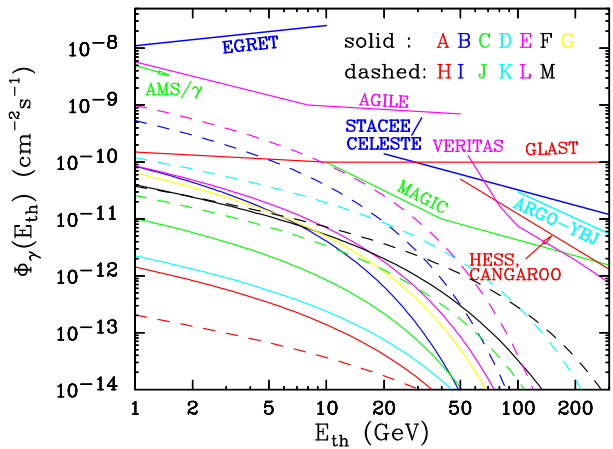}%
\caption{\it The integrated photon flux $\Phi_\gamma(\ethr)$ as a
function of photon energy threshold $\ethr$ for photons produced by
relic annihilations in the galactic center. A moderate halo parameter
$\bar{J} = 500$ is assumed~\cite{Bergstrom:1998fj}.  Point source flux
sensitivities for various gamma ray detectors are also shown. }
\label{fig:photon_spectra}
\end{figure}

We have computed the integrated photon flux $\Phi_\gamma(\ethr)$ in
the direction of the galactic center following the procedure 
of~\cite{Feng:2001zu}. Our results for each of the benchmark points
are presented in Fig.~\ref{fig:photon_spectra}.  Estimates for point
source flux sensitivities of several gamma ray detectors, both current
and planned, are also shown.  The space-based detectors EGRET,
AMS/$\gamma$ and GLAST can detect soft photons, but are limited in
flux sensitivity by their small effective areas.  Ground-based
telescopes, such as MAGIC, HESS, CANGAROO and VERITAS, are much larger
and so sensitive to lower fluxes, but are limited by higher energy
thresholds.  These sensitivities are not strictly valid for
observations of the galactic center.  Nevertheless, they provide rough
guidelines for what sensitivities may be expected in coming years.
For a discussion of these estimates, their derivation, and references
to the original literature, see~\cite{Feng:2001zu}.

Fig.~\ref{fig:photon_spectra} shows that space-based detectors offer
good prospects for detecting a photon signal, while ground-based
telescopes have a relatively limited reach.  GLAST appears to be
particularly promising, with points I and L giving observable signals.
Recall, however, that all predicted fluxes scale linearly with
$\bar{J}$.  For isothermal halo density profiles, the fluxes may be
reduced by two orders of magnitude.  On the other hand, for
particularly cuspy halo models, such as those 
in~\cite{Navarro:1996iw}, all fluxes may be enhanced by two orders
of magnitude, leading to detectable signals in GLAST for almost all
points, and at MAGIC for the majority of benchmark points.


Relic neutralino annihilations in the galactic halo \cite{ss} may also be
detected through positron excesses in space-based and balloon
experiments~\cite{Tylka:1989xj,Moskalenko:1999sb}.  To estimate the
observability of a positron excess, we followed the procedure
advocated in~\cite{Feng:2001zu}. For each benchmark spectrum, we
first find the positron energy $\eopt$ at which the positron signal to
background ratio $S/B$ is maximized.  For detection, we then require
that $S/B$ at $\eopt$ be above some value.  The sensitivities of a
variety of experiments have been estimated in~\cite{Feng:2001zu}.
Among these experiments, the most promising is AMS~\cite{AMS}, the
anti-matter detector to be placed on the International Space Station.
AMS will detect unprecedented numbers of positrons in a wide energy
range.  We estimate that a 1\% excess in an fairly narrow energy bin,
as is characteristic of the neutralino signal, will be statistically
significant. Unfortunately, our study~\cite{newpaper} showed that for
all benchmark points the expected positron signals are below the AMS
sensitivity. However, one should be aware that as with the photon
signal, positron rates are sensitive to the halo model assumed; for
clumpy halos~\cite{Silk:1992bh}, the rate may be enhanced by orders of
magnitude~\cite{Moskalenko:1999sb}.


In conclusion, we have provided indicative estimates of the dark
matter detection rates that could be expected for the benchmark
supersymmetric scenarios proposed in~\cite{Battaglia:2001zp}. We
emphasize that, in addition to the supersymmetric model dependences of
these calculations, there are important astrophysical
uncertainties. These include the overall halo density, the possibility
that it may be enhanced in the solar system, its cuspiness near the
galactic center, and its clumpiness elsewhere. For these reasons, our
conclusions about the relative ease with which different models may be
detected using the same signature may be more reliable than the
absolute strengths of different signatures. Nevertheless, our
estimates do indicate that there may be good prospects for
astrophysical detection of quite a large number of the benchmark
scenarios~\cite{newpaper}.

In particular, four benchmark points (I, B, E and L) may be detected
through spin-independent elastic scattering of relic particles using
the projected GENIUS~\cite{GENIUS} detector, with models G, F and C
not far from the likely threshold of detectability. The indirect
detection of muons generated by high-energy neutrinos due to
annihilations inside the Sun should be most easily detectable with the
proposed IceCube~\cite{IceCube} detector in models E, F and B,
followed by models I, G, L and C, which are near the boundary of
sensititvity. The best prospects for detecting photons from
annihilations in the galactic center (for models L and I) are offered
by the GLAST satellite, with its relatively low threshold. However,
there may also be prospects for ground-based experiments such as MAGIC
if the halo is cuspier at the galactic center than we have assumed.

It was previously noted~\cite{Battaglia:2001zp} that the more $g_\mu -
2$-friendly models (I, L, B, G, C and J) offered good prospects for
detecting several supersymmetric particles at the LHC and/or a 1~TeV
linear $e^+ e^-$ collider. Most of these models also exhibit good
prospects for dark matter detection, with the exception of model
J. Among the less $g_\mu - 2$-friendly models, we note that the focus
points E and F offer good astrophysical prospects, demonstrating the
complementarity of collider and astrophysics searches. This is
particularly interesting in the case of focus-point model F, which is
very challenging for colliders.


\begin{acknowledgments}
The work of J.L.F. was supported in part by the US Department of
Energy under cooperative research agreement DF--FC02--94ER40818.  The
work of K.A.O. was supported partly by DOE grant
DE--FG02--94ER--40823.
\end{acknowledgments}


\end{document}